# A dimensionally split Cartesian cut cell method for the compressible Navier-Stokes equations

Nandan Gokhale[a,*], Nikos Nikiforakis[a], Rupert Klein[b]

[a]*Laboratory for Scientific Computing, Cavendish Laboratory, University of Cambridge, Cambridge, CB3 0HE, UK*
[b]*Institut für Mathematik, FB Mathematik und Informatik, Freie Universität Berlin, Arnimallee 6, 14195 Berlin, Germany*

## Abstract

We present a dimensionally split method for computing solutions to the compressible Navier-Stokes equations on Cartesian cut cell meshes. The method is globally second order accurate in the $L_1$ norm, fully conservative, and allows the use of time steps determined by the regular grid spacing. We provide a description of the three-dimensional implementation of the method and evaluate its numerical performance by computing solutions to a number of test problems ranging from the nearly incompressible to the highly compressible flow regimes. All the computed results show good agreement with reference results from theory, experiment and previous numerical studies. To the best of our knowledge, this is the first presentation of a dimensionally split cut cell method for the compressible Navier-Stokes equations in the literature.

## 1. Introduction

The Cartesian cut cell mesh generation procedure involves computationally 'cutting out' the geometry from a background Cartesian grid to produce a resulting mesh with a sharp representation of the interface. The procedure is attractive since it allows for rapid, automatic mesh generation for complex geometries while retaining the computational conveniences offered by the use of Cartesian grids. Furthermore, cut cell methods, unlike other types of immersed boundary methods [1], are designed to be fully conservative (see Berger [2] for a comprehensive recent overview of cut cell methods). However, the resulting mesh can contain cut cells of arbitrarily small volume fraction at the interface which impose severe constraints on the time step for explicit numerical schemes.

Since the early 1980s, a number of ways have been presented in the literature to deal with this 'small cell problem' in the context of the Euler equations [3, 4, 5, 6, 7, 8]. Over the past decade, most of these techniques have been extended for solving the compressible Navier-Stokes equations. Hartmann et al. [9] have used cell linking (which is related to the intuitive concept of cell merging) to develop a three-dimensional cut cell method that is implemented in an adaptive octree grid. Using the static boundary cut cell formulation of Hartmann et al. [9], Schneiders et al. [10, 11] have successfully developed a method to compute moving boundary problems in 3D by introducing an interpolation routine and flux redistribution step. Berger et al. [12] use pseudo-time stepping and a multigrid approach to compute steady state solutions for the 2D Reynolds-averaged Navier-Stokes (RANS) equations. In a subsequent publication [13], they describe a cut cell implementation of a novel ODE-based wall model which, unlike conventional equilibrium wall functions, has the advantage that it can be applied further away from the interface in the wake region of a turbulent boundary layer. Graves et al. [14] extend the 'flux redistribution' technique of Colella et al. [7] for small cell stability to develop a second order accurate method. Another high order discretisation was demonstrated by Muralidharan and Menon [15], who extended the flux mixing technique of Hu et al. [8] to develop a third order accurate scheme.

*Corresponding author
 *Email addresses:* nbg22@cam.ac.uk (Nandan Gokhale), nn10005@cam.ac.uk (Nikos Nikiforakis),
rupert.klein@math.fu-berlin.de (Rupert Klein)

The aforementioned techniques are all implemented in an unsplit fashion. We are particularly interested in adopting a dimensionally split approach which is a convenient way to extend one-dimensional methods to solve multi-dimensional problems. In that context, Gokhale, Nikiforakis and Klein [16] recently presented a simple dimensionally split cut cell method for hyperbolic conservation laws using a 'Local Proportional Flux Stabilisation' (LPFS) approach and demonstrated its performance through the computation of solutions to a number of challenging problems for the Euler equations. The LPFS method is an improvement on the original split method of Klein, Bates and Nikiforakis (KBN) [17]. Although both methods are first order accurate at the interface, LPFS was shown to produce more accurate solutions near boundaries for hyperbolic problems, and it allows the use of larger Courant numbers [16]. In this paper, we combine and extend the LPFS and KBN methods to solve compressible Navier-Stokes problems involving rigid embedded boundaries. To the best of our knowledge, this is the first presentation of a dimensionally split cut cell method for the compressible Navier-Stokes equations in the literature.

The rest of this paper is organised as follows. In Section 2, we outline the governing equations and solution framework that we use. A description of the cut cell geometric parameters required by the method is provided in Section 3. In Section 4, we describe the numerical method in detail. In Section 5, we present numerical solutions for a number of multi-dimensional test problems to demonstrate the performance of the method. Finally, conclusions and areas for future work are provided in Section 6.

**2. Governing equations and solution framework**

The compressible Navier-Stokes equations are

$$\begin{aligned}
\partial_t \rho + \nabla \cdot (\rho \mathbf{u}) &= 0, \\
\partial_t (\rho \mathbf{u}) + \nabla \cdot (\rho \mathbf{u} \otimes \mathbf{u} + pI) &= \nabla \cdot \sigma, \\
\partial_t E + \nabla \cdot [(E + p)\mathbf{u}] &= \nabla \cdot (\sigma \mathbf{u}) + \nabla \cdot (\xi(\nabla T)),
\end{aligned} \quad (1)$$

where $\rho$ is density, $\mathbf{u}$ is velocity, $p$ is pressure, $I$ is the identity matrix, $\xi$ is thermal conductivity and $T$ is temperature. We assume the fluid under consideration is Newtonian such that $\sigma$, the stress tensor, is

$$\sigma = \mu(\nabla \mathbf{u} + \nabla \mathbf{u}^T) + \lambda(\nabla \cdot \mathbf{u})I, \quad (2)$$

where $\mu$ is the dynamic viscosity and we use the Stokes' hypothesis $\lambda = -\frac{2}{3}\mu$. $E$ is the total energy per unit volume, given by

$$E = \rho \left(\frac{1}{2}|\mathbf{u}|^2 + e\right), \quad (3)$$

where $e$ is the specific internal energy. To close the system of equations Eq. 1-Eq. 3 we use the ideal gas equation of state

$$e = \frac{p}{\rho(\gamma - 1)}, \quad (4)$$

where $\gamma$, the heat capacity ratio, is assumed to be 1.4.

We defer our description of the procedure for computing the explicit inviscid (Godunov-based) and viscous fluxes till Section 4.1, although it may be noted that the flux stabilisation approach is independent of the particular choice of flux methods used. Hierarchical Adaptive Mesh Refinement (AMR) [18] is used to refine areas of interest such as the cut cell interface, or shock waves, while allowing the use of coarser resolutions elsewhere for the sake of computational efficiency. In this work, we manually specify the extent of the refined region for all subsonic problems. For supersonic problems, dynamic refinement of the region covering the shock is carried out by comparing a density gradient based metric to a user-defined threshold. Multi-dimensional updates are performed using Strang splitting [19] in 2D, and straightforward Godunov splitting [20] in 3D, although Strang splitting could also be used in 3D if time order of accuracy was important for the problem at hand.

As in Dragojlovic et al. [21], we assume that the global time step, $\Delta t$, is restricted by the minimum of the hyperbolic and diffusive time steps

$$\Delta t = \min\left[\Delta t_{\text{hyp}}, \Delta t_{\text{diff}}\right],$$
$$= \min\left[C_{\text{cfl}} \min_{d,i}\left(\frac{\Delta x_{d,i}}{W_{d,i}^{\max}}\right), \min_{d,i}\left(\frac{\Delta x_{d,i}^2}{2\max(\frac{\mu_i}{\rho_i}, \frac{\xi_i}{(\rho c_p)_i})}\right)\right], \quad (5)$$

where $d$ is the index of the coordinate direction, $i$ is the index of a computational cell, and $c_p$ is the specific heat at constant pressure. $\Delta x_{d,i}$ and $W_{d,i}^{\max}$ are the spatial resolution and max wave speed for cell $i$ in the $d$ direction respectively. As in similar previous works [21, 9, 10, 15], we operate the algorithm in the region where the hyperbolic condition is most restrictive, and all simulations in this paper were run using a Courant number of 0.5.

The wave speed for cell $i$ in the $d$ direction, $W_{d,i}$, is computed using the following estimate suggested by Toro [22]:

$$W_{d,i} = |\mathbf{u}_{d,i}| + a_i, \quad (6)$$

where $\mathbf{u}_{d,i}$ is the component of the velocity in cell $i$ in the $d$ direction. $a_i$ is the speed of sound in cell $i$, given by

$$a_i = \sqrt{\frac{\gamma p_i}{\rho_i}}. \quad (7)$$

## 3. Mesh generation

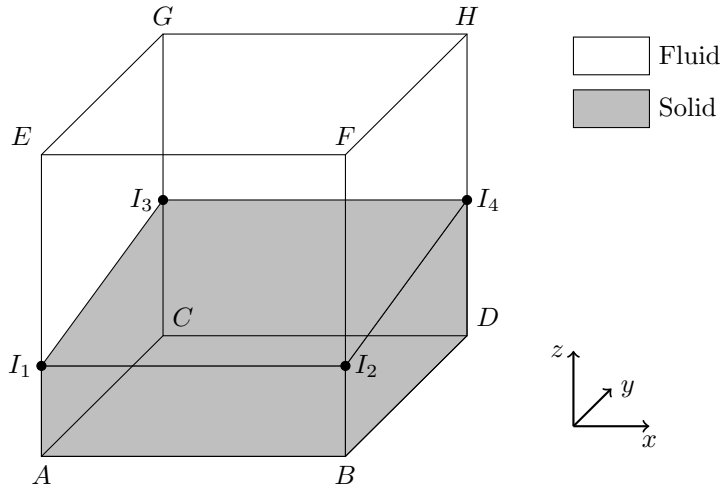

Figure 1: Illustration of a cut cell in 3D.

Consider Fig. 1, which shows a 3D cut cell where the solid-fluid interface intersects the cell at four points $I_1$, $I_2$, $I_3$ and $I_4$. The geometric parameters required by the method are the following:

(i) The face fraction, $\beta \in [0,1]$, of each cell face. This represents the fluid area of the face non-dimensionalised by total cell face area.

(ii) The area, $A^b$, of the reconstructed interface in the cell and $\hat{\mathbf{n}}^b$, the interface unit normal. The superscript $b$ is short for 'boundary'.

(iii) The volume fraction, $\alpha \in [0, 1]$, of the cell. This is the fluid volume of the cell non-dimensionalised by the total cell volume.

(iv) The volumetric centroid, $\mathbf{x}_c$, of the fluid part of the cell.

(v) The interface centroid, $\mathbf{x}_c^b$ of the reconstructed solid interface.

We proceed by treating the interface implicitly as the zero level-set of a signed distance function. The technique of Mauch [23] is used to compute the signed distance function, $\phi(\mathbf{x})$, at the vertices of the cell. As described in [16], this information can be used to work out the coordinates of the intersection points $I_1$-$I_4$ and parameters (i)-(iii) defined above. We avoid repeating the details here for the sake of brevity. It may be noted that our mesh generation procedure is only valid when the intersection of the cell and geometry can be described by a single interface. Sufficient resolution must therefore be used to ensure that all cut cells are 'singly cut'. The signed distance function can be used to deal with 'split cells' created by multiple intersections of the geometry with the cell by using a 'Marching Cubes' [24] based approach as in Gunther et al. [25], but this is beyond the scope of this work. Next, we describe the calculation of parameters (iv) and (v) in two and three dimensions.

In 2D, the fluid part of a cut cell is a polygon. $\mathbf{x}_c = [\mathbf{x}_{c,x}, \mathbf{x}_{c,y}]^T$ can therefore be worked out using the formula for the centroid of a non-self-intersecting polygon with $n$ ordered vertices $(x_1, y_1), ..., (x_n, y_n)$ [26]

$$\mathbf{x}_{c,x} = \frac{1}{6A} \sum_{i=1}^{n} (x_i + x_{i+1})(x_i y_{i+1} - x_{i+1} y_i), \tag{8}$$

$$\mathbf{x}_{c,y} = \frac{1}{6A} \sum_{i=1}^{n} (y_i + y_{i+1})(x_i y_{i+1} - x_{i+1} y_i), \tag{9}$$

where $A$ is the area of the polygon [26]

$$A = \frac{1}{2} \sum_{i=1}^{n} (x_i y_{i+1} - x_{i+1} y_i). \tag{10}$$

Note that in Eq. 8-Eq. 10, the summation index is periodic so that $(n + 1) = 1$.

Since we assume that cut cells are singly cut, $\mathbf{x}_c^b$ in 2D is worked out trivially as the average of the positions of the two intersection points.

In 3D, the fluid part of the cell is a polyhedron. $\mathbf{x}_c$ can therefore be calculated using the formula for the centroid of an arbitrary polyhedron with $N_F$ faces [27]

$$\mathbf{x}_c = \frac{3}{4} \frac{\left[\sum_{i=1}^{N_F} (\mathbf{x_{c,i}} \cdot \mathbf{\hat{n}_i}) \mathbf{x_{c,i}} A_i\right]}{\left[\sum_{i=1}^{N_F} (\mathbf{x_{c,i}} \cdot \mathbf{\hat{n}_i}) A_i\right]}, \tag{11}$$

where $\mathbf{x_{c,i}}$ is the centroid of face $i$ (computed by the appropriate use of Eq. 8-Eq. 10), and $\mathbf{\hat{n}_i}$ and $A_i$ are the outward unit normal and area of face $i$ respectively.

To compute $\mathbf{x}_c^b$ in 3D, we follow the following procedure:

1. Rotate the interface plane with $n$ vertices $(x_1, y_1, z_1), ..., (x_n, y_n, z_n)$ to give a polygon aligned with the $x$-$y$ Cartesian plane. This polygon has vertices $(x_1^{\text{rot}}, y_1^{\text{rot}}, z^{\text{rot}}), ..., (x_n^{\text{rot}}, y_n^{\text{rot}}, z^{\text{rot}})$. Note that because the interface reconstructed from the signed distance function may not be perfectly planar, the $z$ coordinates of the rotated interface may differ slightly from one another. In practice, we set $z^{\text{rot}}$ to be the numerical average of these varying $z$ coordinates.

2. Use Eq. 8-Eq. 10 on the polygon with ordered vertices $(x_1^{\text{rot}}, y_1^{\text{rot}}), ..., (x_n^{\text{rot}}, y_n^{\text{rot}})$ to get $\mathbf{x}_{c,x}^{b,\text{rot}}$ and $\mathbf{x}_{c,y}^{b,\text{rot}}$, which are the $x$ and $y$ coordinates of the interface centroid in the rotated frame, $\mathbf{x}_c^{b,\text{rot}}$. Note that $\mathbf{x}_c^{b,\text{rot}} = [\mathbf{x}_{c,x}^{b,\text{rot}}, \mathbf{x}_{c,y}^{b,\text{rot}}, z^{\text{rot}}]^T$.

3. Rotate $\mathbf{x}_c^{b,\text{rot}}$ back to the original frame to give $\mathbf{x}_c^b$ as required.

# 4. Numerical method

## 4.1. Calculation of explicit fluxes

In this subsection, we describe the procedure we use to compute the explicit inviscid and viscous fluxes. Their stabilisation at the cut cells is described in Section 4.2.

### 4.1.1. Intercell fluxes

In this work, we compute the inviscid intercell fluxes that appear in the divergence terms to the left of the equal signs in Eq. 1 using an exact Riemann solver and the MUSCL-Hancock scheme in conjunction with the van-Leer limiter [22]. This scheme is second order accurate in smooth regions.

The calculation of the viscous fluxes that appear in the divergence terms to the right of the equal signs in Eq. 1 requires the computation of $\nabla \mathbf{u}$ and $\nabla T$ at the cell faces. In Fig. 2, we illustrate our procedure for calculating $(\nabla \phi)_{i-1/2,j}$ for a general scalar $\phi$ at the left face of cell $(i,j)$ in an $x$ dimensional sweep in two dimensions.

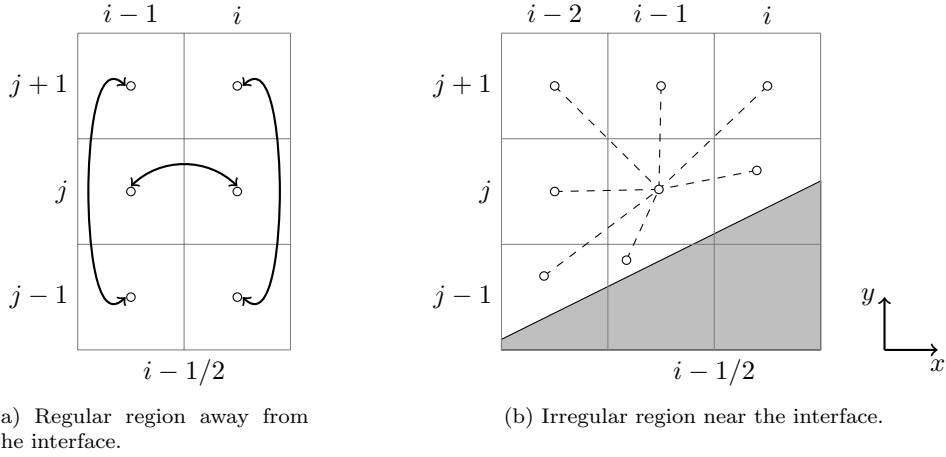

(a) Regular region away from the interface.

(b) Irregular region near the interface.

Figure 2: Illustration of the calculation of viscous face derivatives in irregular and regular regions.

In a regular region away from the interface, the required face derivatives can be worked out using central differencing as illustrated in Fig. 2a. The derivative in the current coordinate direction is calculated to second order accuracy as

$$\left(\frac{\partial \phi}{\partial x}\right)_{i-1/2,j} = \frac{\phi_i - \phi_{i-1}}{\Delta x}. \tag{12}$$

The transverse derivative is computed to second order accuracy as the average of the neighbouring transverse derivatives

$$\left(\frac{\partial \phi}{\partial y}\right)_{i-1/2,j} = \frac{1}{2}\left(\frac{\phi_{i-1,j+1} - \phi_{i-1,j-1}}{2\Delta y} + \frac{\phi_{i,j+1} - \phi_{i,j-1}}{2\Delta y}\right). \tag{13}$$

It may be noted that in 3D, the face derivatives can be calculated analogously on a $2 \times 3 \times 3$ point stencil.

The procedure for irregular regions is illustrated in Fig. 2b. We start by computing the inverse distance weighted least squares gradients $(\nabla \phi)^{\text{LS}}_{i-1,j}$ and $(\nabla \phi)^{\text{LS}}_{i,j}$ at the volumetric centroids of cells $(i-1,j)$ and $(i,j)$ respectively. Note that the stencil for the weighted least squares calculations contains all the regular and cut cells from the 8 (26 in 3D) neighbours of the cell. This is illustrated using dotted lines for cell $(i-1,j)$ in Fig. 2b. $(\nabla \phi)_{i-1/2,j}$ is calculated as

$$(\nabla \phi)_{i-1/2,j} = \frac{(\nabla \phi)^{\text{LS}}_{i-1,j} + (\nabla \phi)^{\text{LS}}_{i,j}}{2}. \tag{14}$$

As with other cut cell discretisations [9, 12], our approximation of the face derivatives in irregular regions is only first order accurate. This does not impact the overall accuracy of the method, however, since the flux stabilisation is also first order accurate at the boundary.

*4.1.2. Boundary fluxes*

To ensure conservation in a dimensionally split scheme, advective boundary fluxes have to be treated differently to the pressure and diffusive fluxes. We describe the treatment for the inviscid Euler equations in [16]. Essentially, the advective boundary flux in a cell needs to be evaluated using a 'reference' boundary state computed by solving a wall normal Riemann problem. The reference state should be computed before the start of a time step and kept constant for the duration of the time step. The velocity resulting from the wall Riemann problem is tangential to the wall. As a consequence, the sum of advective fluxes across the interface accummulated over a full cycle of split steps cancels exactly, thus ensuring zero net mass flux across the wall. The boundary pressure, on the other hand, is to be updated in between sweeps for the sake of accuracy. For the Navier-Stokes equations, we need to compute also the stress tensor at the boundary and update it in between sweeps in order to compute accurate viscous momentum boundary fluxes. Since we consider only static (no-slip), adiabatic boundaries in this work, note that there are no viscous heating or thermal conductivity boundary fluxes to consider for the energy equation.

To compute $\nabla \mathbf{u}$ at the boundary, we follow the process illustrated for cut cell $(i,j)$ in Fig. 3. Note that in the rest of this section, subscripts specified with greek letters are assumed to range from 1 to $n_d$, the number of dimensions, and repeated indices of that kind imply the use of the Einstein summation convention. Let $x_\mu$ represent a Cartesian coordinate system, and $\hat{x}_\mu$ represent an orthonormal coordinate system with a unit vector pointing normal to the cell interface, and unit vector(s) in the interface tangential plane.

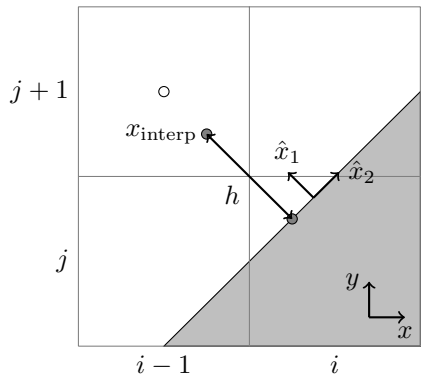

Figure 3: Illustration of the procedure used to compute $\nabla \mathbf{u}$ at the boundary of cell $(i,j)$. $u_\mu^*$ at $x_{\text{interp}}$ is reconstructed from the weighted least squares gradient computed at the nearest cell volumetric centroid (filled white circle).

We start by computing the normal boundary derivatives for each velocity component $u_\mu$ as

$$\left(\frac{\partial u_\mu}{\partial \hat{x}_1}\right) = \frac{u_\mu^* - u_\mu^b}{h} = \frac{u_\mu^*}{h}, \tag{15}$$

where $u_\mu^b$, the value of the velocity component at the boundary is 0 as required by the no-slip boundary condition. $u_\mu^*$ is the value of the velocity component at the interpolation point $x_{\text{interp}}$ which is located at a distance of $h$ from the interface centroid in the normal direction. We reconstruct the value of $u_\mu^*$ using a weighted least squares gradient computed at the cell volumetric centroid closest to $x_{\text{interp}}$. $h$ is calculated as in Meyer et al. [28] using

$$h = 0.5\sqrt{\hat{\mathbf{n}}_\nu^b \Delta x_\nu^2}, \tag{16}$$

where $\Delta x_\nu$ is the spatial resolution in the $\nu$ coordinate direction. Note that when using a uniform mesh spacing of $\Delta x$ in all coordinate directions, $h = 0.5\Delta x$.

With the normal derivatives calculated, and the tangential boundary derivatives known to be 0 because of no-slip and no space dependent wall motion, we can construct the tensor $\partial u_\mu / \partial \hat{x}_\xi$ at the boundary. However, we want to compute the tensor $\nabla \mathbf{u}$ which has Cartesian components $\partial u_\mu / \partial x_\nu$ that are given by

$$\frac{\partial u_\mu}{\partial x_\nu} = \frac{\partial \hat{x}_\xi}{\partial x_\nu} \frac{\partial u_\mu}{\partial \hat{x}_\xi}. \tag{17}$$

$\partial \hat{x}_\xi / \partial x_\nu$ can be represented by a matrix whose columns are the unit vectors of the $\hat{x}_\mu$ coordinate system. The normal vector $\hat{\mathbf{n}}^b$ is already known from the information provided by the signed distance function (see Section 3), while the vectors spanning the tangential plane are calculated from $\hat{\mathbf{n}}^b$ using a Gram-Schmidt orthogonalisation process. Eq. 17 can then be used to compute $\nabla \mathbf{u}$, and hence the stress tensor $\sigma$ at the boundary as required to compute the momentum diffusion boundary fluxes.

Finally, although beyond the scope of this work, it may be noted that if needed, it is possible to introduce the use of a simple algebraic or ODE-based turbulent wall function into the above approach [12, 29, 30]. The reconstructed state at $x_\text{interp}$ and the known state at the wall interface centroid can be used as boundary conditions to calculate the wall shear stress.

*4.2. Flux stabilisation*

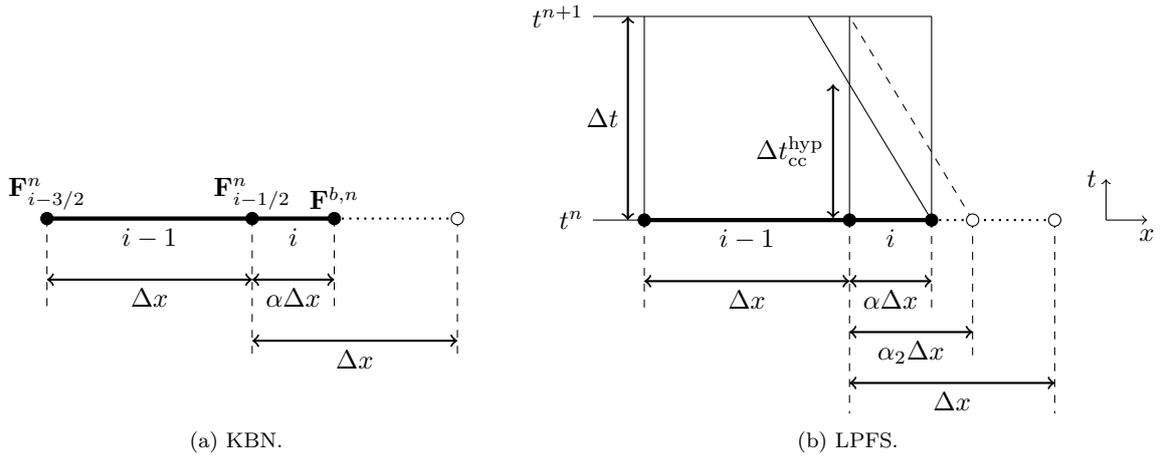

(a) KBN.  (b) LPFS.

Figure 4: Illustration of the KBN and LPFS flux stabilisation procedures for a boundary cut cell neighbouring a regular cell in 1D.

In this subsection, we describe the combination and extension of the KBN and LPFS methods for the Navier-Stokes equations. Fig. 4a and Fig. 4b illustrate the one-dimensional flux stabilisation procedures used in both approaches for a boundary cut cell $i$ neighbouring a regular cell.

Let $\mathbf{U}_i^n$ represent the conserved variable state vector for cell $i$ at time level $n$, and let $\mathbf{F}_{i\pm i/2}^n$ represent the explicit numerical fluxes (inviscid and viscous) computed at its ends. As described in [16, 17], the stabilised KBN flux can be derived to be

$$\mathbf{F}_{i-1/2}^{\text{KBN},n} = \mathbf{F}^{b,n} + \alpha(\mathbf{F}_{i-i/2}^n - \mathbf{F}^{b,n}), \tag{18}$$

where we note that the expression is consistent with respect to the natural limits of the grid so that as $\alpha \to 1$, $\mathbf{F}_{i-1/2}^{\text{KBN},n} \to \mathbf{F}_{i-i/2}^n$, and as $\alpha \to 0$, $\mathbf{F}_{i-1/2}^{\text{KBN},n} \to \mathbf{F}^{b,n}$.

The KBN flux Eq. 18 uses only the geometric parameter $\alpha$ to derive a stable flux. The LPFS approach combines both geometric and wave speed information to derive a flux that was shown in [16] to produce more accurate boundary solutions for hyperbolic problems. Consider Fig. 4b, which shows the boundary cut cell neighbouring the regular cell in the $x$-$t$ plane for one time step. $\Delta t$ is the global stable hyperbolic time step which is determined in part by the fastest wave speed in the domain, $W_{\max}$ (see Eq. 5). For the configuration of Fig. 4b, we illustrate the 'small cell problem' at the cut cell as being caused by the left-going wave from the solution of the boundary Riemann problem. Stability would therefore require the use of the smaller $\Delta t_{\mathrm{cc}}^{\mathrm{hyp}}$

$$\Delta t_{\mathrm{cc}}^{\mathrm{hyp}} = C_{\mathrm{cfl}} \frac{\alpha \Delta x}{W_i}, \tag{19}$$

where $W_i$ is the wave speed for the cut cell.

In the LPFS approach, the basic idea is to use the explicit flux $\mathbf{F}_{i-i/2}^n$ for the part of the time step for which it is stable, $\Delta t_{\mathrm{cc}}^{\mathrm{hyp}}$, and to employ a different flux which can maintain stability for the duration $(\Delta t_{\mathrm{cc}}^{\mathrm{hyp}}, \Delta t]$. This gives the LPFS method an inherent advantage in regions of low velocity - although $\alpha < 1$ depresses $\Delta t_{\mathrm{cc}}^{\mathrm{hyp}}$ relative to $\Delta t$, part of the reduction is offset if $W_i < W_{\max}$, an effect which is most pronounced in regions of low velocity near a stagnation point or in the boundary layer. For larger cut cells, the ratio $\Delta t_{\mathrm{cc}}^{\mathrm{hyp}}/\Delta t$ can be greater than 1 so that no flux stabilisation is required.

As described in [16], a suitable LPFS flux is

$$\mathbf{F}_{i-1/2}^{\mathrm{LPFS},n} = \frac{\Delta t_{\mathrm{cc}}^{\mathrm{hyp}}}{\Delta t} \mathbf{F}_{i-1/2}^n + \left(1 - \frac{\Delta t_{\mathrm{cc}}^{\mathrm{hyp}}}{\Delta t}\right) \mathbf{F}_{i-1/2}^{\mathrm{KBN,mod},n}, \tag{20}$$

where

$$\mathbf{F}_{i-1/2}^{\mathrm{KBN,mod},n} = \mathbf{F}^{b,n} + \frac{\alpha}{\alpha_2}(\mathbf{F}_{i-i/2}^n - \mathbf{F}^{b,n}), \tag{21}$$

and

$$\frac{\alpha}{\alpha_2} = \frac{\Delta t_{\mathrm{cc}}^{\mathrm{hyp}}}{\Delta t} = \epsilon \frac{\alpha W_{\max}}{W_i}. \tag{22}$$

The 'wave speeds uncertainty' parameter $\epsilon \in [0,1]$ is introduced to account for any errors arising from the use of Eq. 6 to estimate the cut cell wave speeds. We found setting $\epsilon$ to 0.8 to be a robust choice for the wide range of problems tackled in this paper.

The derivation of the LPFS flux Eq. 20 implicitly assumes that the cut cell time step is limited by the local hyperbolic time step $\Delta t_{\mathrm{cc}}^{\mathrm{hyp}}$. Although the global time step for all problems in this work is indeed limited by the hyperbolic time step, it is possible for the local time step at some cut cells to be limited by the local diffusion time step

$$\Delta t_{\mathrm{cc}}^{\mathrm{diff}} = \frac{(\alpha \Delta x)^2}{2 \max(\frac{\mu_i}{\rho_i}, \frac{\xi_i}{(\rho c_p)_i})}, \tag{23}$$

in which case one can use the KBN stabilisation Eq. 18 but not the LPFS stabilisation.

For a completely stable discretisation, then, we compare the relative magnitudes of $\Delta t_{\mathrm{cc}}^{\mathrm{hyp}}$ and $\Delta t_{\mathrm{cc}}^{\mathrm{diff}}$ before deciding how to stabilise the flux. For the usual case $\Delta t_{\mathrm{cc}}^{\mathrm{hyp}} \leq \Delta t_{\mathrm{cc}}^{\mathrm{diff}}$, we use the LPFS flux Eq. 20 at the cut cell cell. If $\Delta t_{\mathrm{cc}}^{\mathrm{diff}} < \Delta t_{\mathrm{cc}}^{\mathrm{hyp}}$, on the other hand, we use the KBN flux Eq. 18.

### 4.3. Multi-dimensional extension

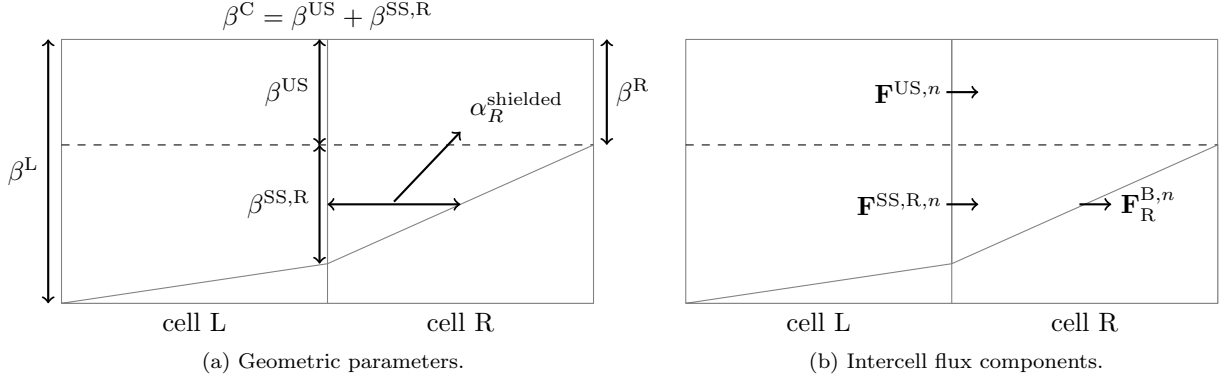

Figure 5: Illustration of the parameters used in the flux stabilisation process for multi-dimensional simulations.

As described in detail in [17, 16], the extension of the 1D LPFS and KBN flux stabilisation approaches to multiple dimensions requires attention to be given to the irregular nature of the cut cells. With reference to Fig. 5, we briefly summarise the procedure for computing the intercell flux between neighbouring cells for an arbitrary dimensional sweep in 2D. The process is exactly the same in 3D.

The basic idea is to divide the interface into 'unshielded' (US) regions which do not 'face' the boundary, and 'shielded' regions which are 'covered' by the boundary in the current sweep direction. To minimise the diffusive impact of the first order flux stabilisation, the aim is to deviate as little as possible from the standard explicit update for regular cells. The flux on the unshielded region does not call for stabilisation so the regular explicit flux (calculated as per Section 4.1.1) is applied there. The stabilised flux is applied only on the shielded regions. In Fig. 5, part of the interface is 'singly-shielded from the right' (SS,R) and so the flux there would be stabilised, with the place of $\alpha$ in Eq. 18 or Eq. 20 being taken by $\alpha_R^{\text{shielded}}$, which is the average distance from the cell face to the boundary in the current coordinate direction, non-dimensionalised by the corresponding regular cell spacing. The boundary flux $\mathbf{F}_R^{B,n}$ is calculated as per Section 4.1.2. Note that all the geometric parameters of Fig. 5a can be computed from the geometric information described in Section 3. The details are left out for the sake of brevity.

The single modified flux $\mathbf{F}_C^{\text{modified},n}$ to be applied at the cell face, then, is taken as an area-weighted sum of the individual components:

$$\mathbf{F}_C^{\text{modified},n} = \frac{1}{\beta^C}\left[\beta^{\text{US}}\mathbf{F}^{\text{US},n} + \beta^{\text{SS,R}}\mathbf{F}^{\text{SS,R},n}\right]. \tag{24}$$

With the modified fluxes computed, the multi-dimensional update for a cell $(i,j)$ using straightforward Godunov splitting, for example, would be given by

$$\mathbf{U}_{i,j}^{n+1/2} = \mathbf{U}_{i,j}^n + \frac{\Delta t}{\alpha_{i,j}\Delta x}\left[\beta_{i-1/2,j}\mathbf{F}_{i-1/2,j}^{\text{modified},n} - \beta_{i+1/2,j}\mathbf{F}_{i+1/2,j}^{\text{modified},n} - \left(\beta_{i-1/2,j} - \beta_{i+1/2,j}\right)\mathbf{F}_{i,j}^{B,n}\right],$$

$$\mathbf{U}_{i,j}^{n+1} = \mathbf{U}_{i,j}^{n+1/2} + \frac{\Delta t}{\alpha_{i,j}\Delta y}\left[\beta_{i,j-1/2}\mathbf{G}_{i,j-1/2}^{\text{modified},n} - \beta_{i,j+1/2}\mathbf{G}_{i,j+1/2}^{\text{modified},n} - \left(\beta_{i,j-1/2} - \beta_{i,j+1/2}\right)\mathbf{G}_{i,j}^{B,n}\right], \tag{25}$$

where we use $\mathbf{F}$ and $\mathbf{G}$ to denote fluxes acting in the $x$ and $y$ directions respectively. A 3D simulation would involve one more sweep using fluxes $\mathbf{H}$ acting in the $z$ direction.

For concave geometries, it is quite possible for part of the interface to be shielded by the boundary from both the left and right sides. Devising a completely stable flux to be applied in these 'doubly-shielded' regions is a challenging research problem. A simple practical solution is to use the 'mixing flux' and 'conservative correction' procedures as described in [16].

# 5. Results

## 5.1. $Re = 20$ lid-driven cavity problem

The lid-driven cavity problem from Kirkpatrick et al. [31] was used to verify the accuracy of the numerical discretisation. Fig. 6a illustrates the simulation set-up. The left, right and bottom domain edges are static no-slip boundaries. The top boundary is also no-slip, but with a parabolic $x$ velocity profile $u_{\text{lid}}$ which varies from 0 at the boundary edges to $u_{\text{lid}}^{\max}$ at the centre. $u_{\text{lid}}^{\max}$ was set to correspond to $M = 0.1$. The Reynolds number of the flow based on the lid length $L$ and $u_{\text{lid}}^{\max}$ is 20.

Fig. 6a shows the contours of normalised $x$ velocity for the 'reference' solution computed on a fine $400 \times 400$ grid, on which the minimum encountered cut cell volume fraction was $2.5 \times 10^{-5}$. The simulation was performed at five coarser resolutions, with the errors for these runs computed relative to the reference solution. The dashed box in Fig. 6a shows the region used for the error computations. In [16], we showed that for a scheme which is first order accurate at the cut cells and second order accurate elsewhere, the $L_p$ norm of the global solution error converges as $\mathcal{O}(\Delta x^{\frac{p+1}{p}})$. As seen in Fig. 6b, the computed solution converges with first order at the cut cells, while the global error measured by the $L_1$ norm converges with second order accuracy as expected.

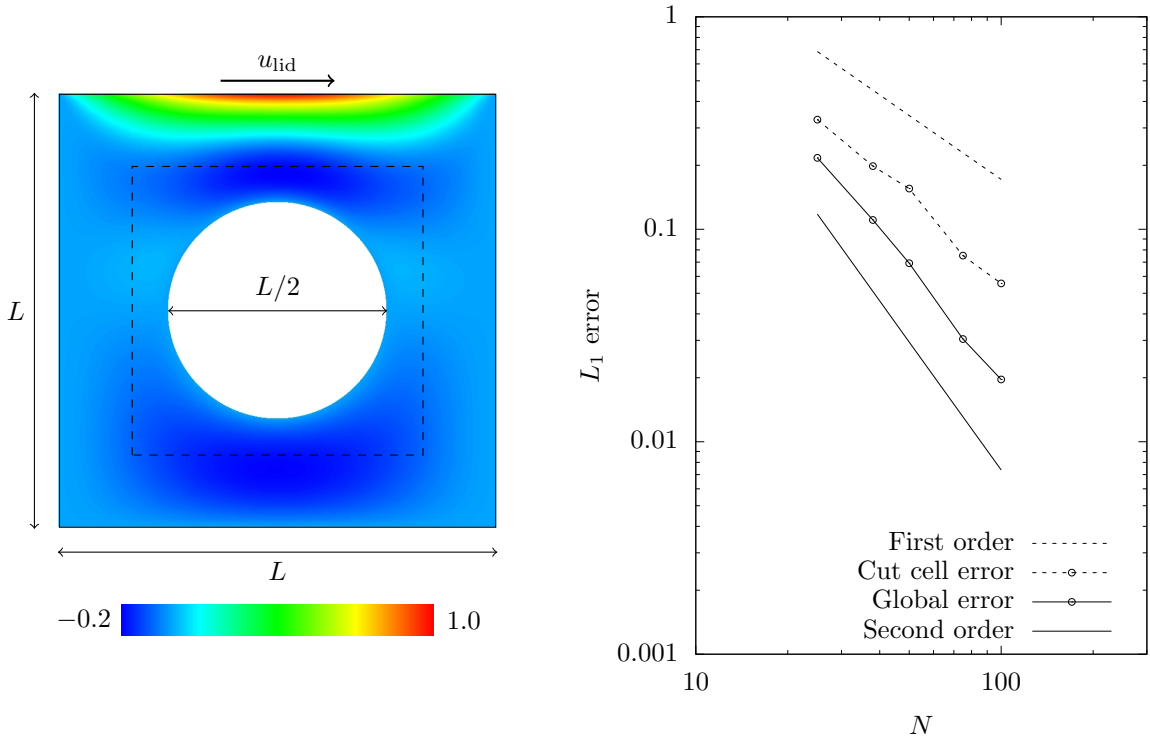

(a) $u/u_{\text{lid}}^{\max}$ contours for the $400 \times 400$ cells reference solution.

(b) Convergence of cut cell and global $L_1$ error norms for velocity magnitude. $N$ is the number of cells along one coordinate direction.

Figure 6: Velocity results for the $Re = 20$ lid-driven cavity problem.

## 5.2. Laminar flat plate boundary layer

This test involves the computation of a two-dimensional flat plate boundary layer for $M_\infty = 0.2$ and $Re_L = 30000$. We run the test in configurations with the plate both coordinate aligned and non-aligned as illustrated in Fig. 7. AMR is used to refine the no-slip region of the plate. Note that a uniform velocity profile is specified at the inflow boundary so that the boundary layer develops on the no-slip part of the plate.

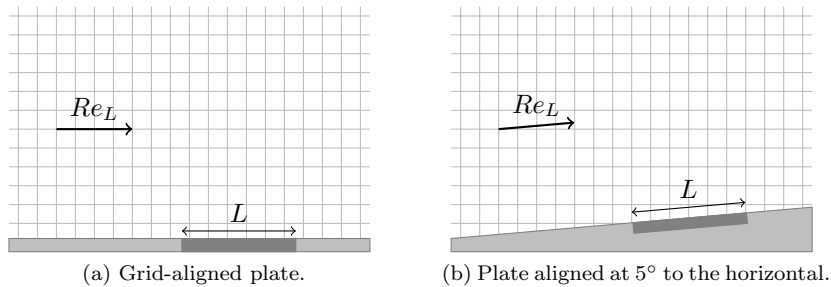

(a) Grid-aligned plate.   (b) Plate aligned at 5° to the horizontal.

Figure 7: Grid aligned and non-aligned configurations for the laminar flat plate boundary layer problem. The shaded region of length $L$ is the no-slip part of the plate. The remainder of the plate is specified as a slip boundary.

For the grid-aligned configuration, the first row of finite volumes adjacent to the plate are all cut cells with the same volume fraction. Fig. 8a shows a comparison of the computed boundary layer profile at the centre of the plate for a series of 4 simulations with progressively smaller cut cell volume fractions. Four levels of AMR are employed such that on the finest level, there are roughly 30 cells resolving the 99% boundary layer thickness $\delta_{99}$. All the computed profiles show good agreement with the theoretical Blasius solution, which is a similarity solution for a steady, two-dimensional laminar boundary layer forming over semi-infinite plate in an incompressible flow [32].

We use twice the resolution for the non-aligned plate configuration such that there are roughly 60 cells resolving the $\delta_{99}$ thickness at the centre of the plate. Like Graves et al. [14], we compare the computed boundary layer velocity profiles along 'wall normal rays' emanating from every cut cell in the range $5000 \leq Re_x \leq 15000$, which covers cut cells of varying shapes and volume fractions. As seen in Fig. 8b, the solutions overlay well and show good agreement with the theoretical Blasius solution.

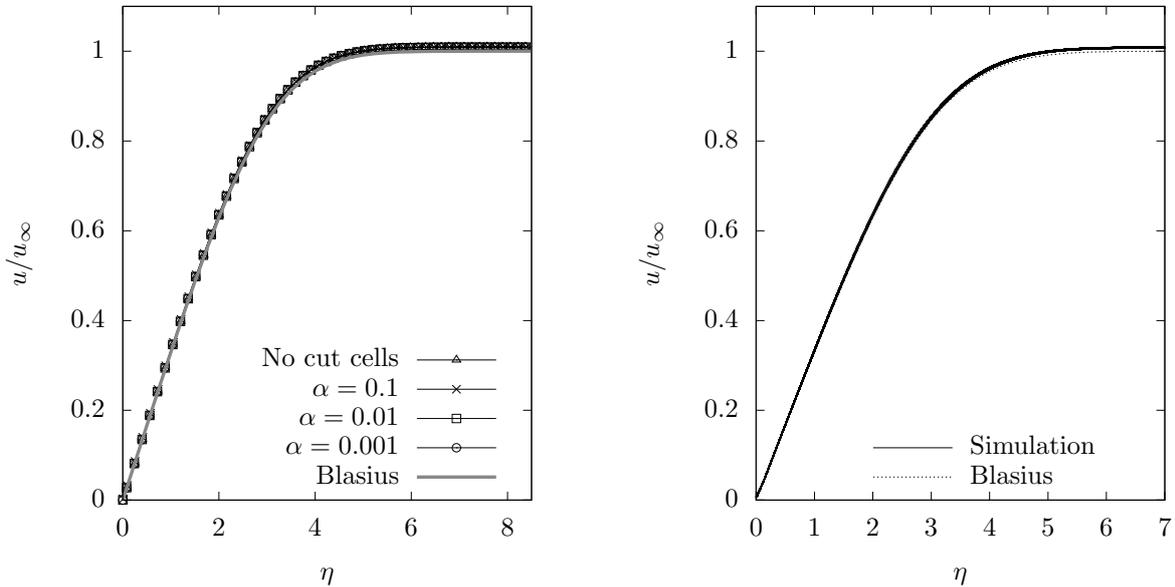

(a) Boundary layer velocity profiles at $Re_x = 15000$ for varying cut cell volume fractions of the lowest row for the grid-aligned configuration.

(b) Velocity profiles along wall-normal rays emanating from every cut cell in the range $5000 \leq Re_x \leq 15000$ for the slanted wall configuration with a wall angle of $5°$.

Figure 8: Computed boundary layer profiles for the (a) co-ordinate aligned, and (b) non-aligned, laminar flat plate boundary layer problem. Note that $\eta = y\sqrt{\frac{u}{\nu x}}$.

*5.3. Flow over a circular cylinder*

The next problem considered is that of two-dimensional flow over a circular cylinder at $Re = 40$ and $Re = 100$. Large amounts of experimental and numerical results exist in the literature for both cases. At $Re = 40$, the eddies behind the cylinder are attached and we can evaluate the accuracy and convergence rate of the computed surface solutions. At $Re = 100$, the wake is unstable and we can also compute the frequency of the vortex shedding process.

The simulation set-up is illustrated in Fig. 9a. The left boundary is subsonic inflow while subsonic outflow boundary conditions are specified on the other domain edges. As illustrated in Fig. 9b, AMR is used to resolve the near wall solution and cylinder wake. $M_\infty$ is set to 0.1.

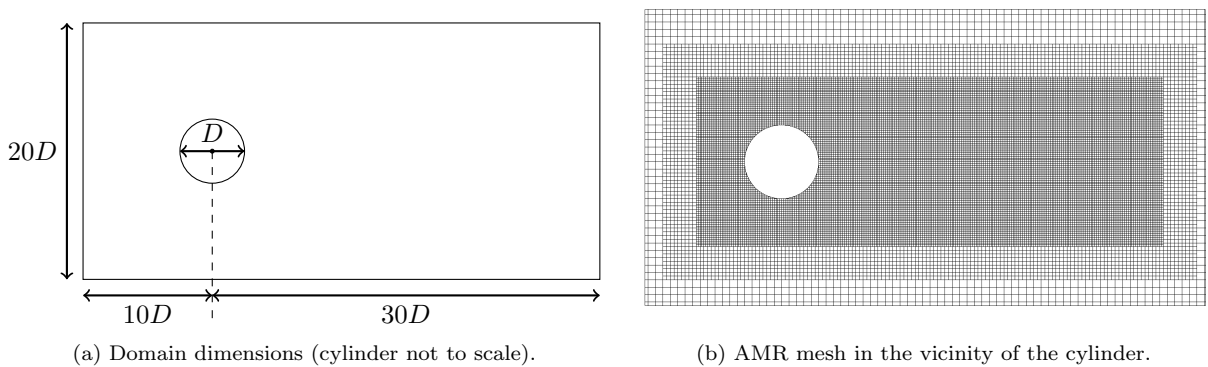

(a) Domain dimensions (cylinder not to scale).

(b) AMR mesh in the vicinity of the cylinder.

Figure 9: Domain dimensions and an illustration of the Adaptive Mesh Refinement used for the simulations of the flow over a circular cylinder.

The simulations for both Reynolds numbers are run at 4 resolutions which are specified such that there are 40, 60, 80 and 160 cells respectively resolving the cylinder diameter $D$ on the finest AMR level. At the finest resolution, the minimum encountered cut cell volume fraction is $6.7 \times 10^{-5}$. The results from the finest resolution simulations are treated as the reference solutions to which errors from the coarser simulations are compared.

As shown in Table 1, the computed drag coefficient from the reference solution for the $Re = 40$ case is in very good agreement with previous experimental and numerical studies.

Table 1: Comparison of the computed drag coefficient with previous experimental and numerical studies for the $Re = 40$ cylinder flow problem.

| Study | $C_D$ |
|---|---|
| Present work (resolution: $D/\Delta x = 160$) | 1.57 |
| Tritton [33] (experiment) | 1.57 |
| Tseng and Ferziger [34] (simulation, resolution: $\pi D/\Delta x \approx 72$) | 1.53 |
| Meyer et al. [28] (simulation, resolution: $D/\Delta x = 72$) | 1.56 |

Fig. 10a shows the computed pressure distribution over the cylinder for the $Re = 40$ case. The results compare well with experimental measurements from Grove et al. [35] over the whole of the cylinder. Furthermore, as shown in Fig. 10b the computed drag coefficient shows the expected first order convergence rate to the reference solution.

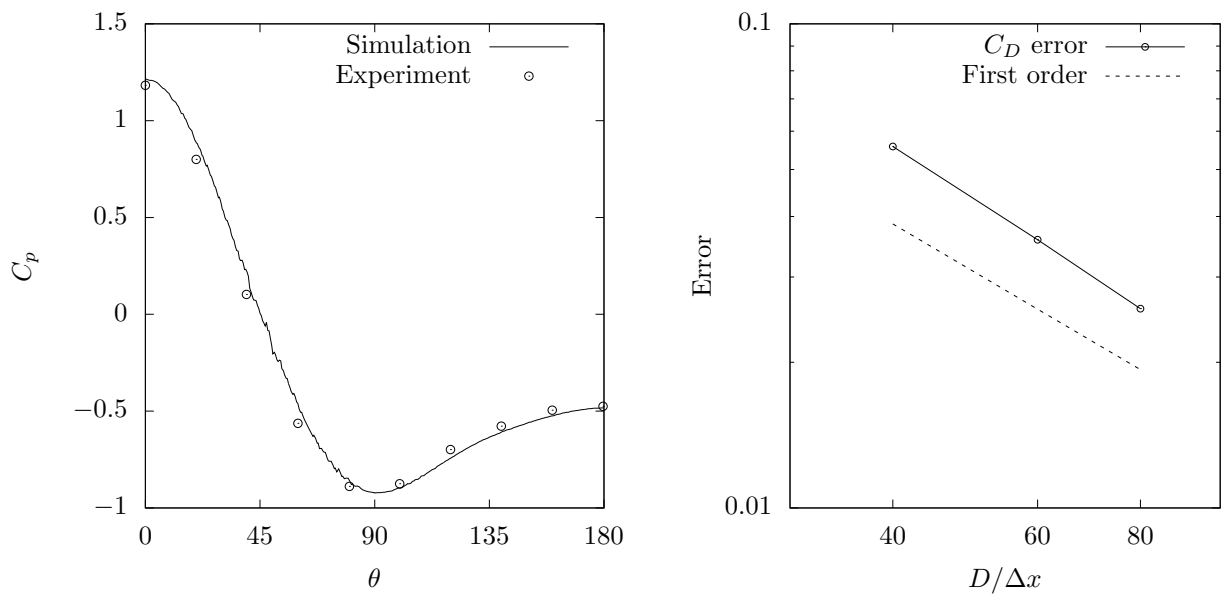

(a) Comparison of the numerical and experimental [35] surface pressure distributions.

(b) Convergence plot of $C_D$ computation.

Figure 10: Results of surface pressure distribution and drag coefficient convergence for the $Re = 40$ cylinder flow problem.

At $Re = 100$, the simulation produces a periodic vortex shedding flow pattern (Fig. 11a shows instantaneous computed vorticity contours in the wake of the cylinder). Table 2 compares the computed time-averaged drag coefficient $\overline{C_D}$ and Strouhal number $St$ with previous experimental and numerical studies. The frequency of the vortex shedding is calculated from the Fourier Transform of the solution for $y$ velocity at a point in the middle of the wake located a distance of $1.5D$ behind the cylinder. The simula-tion results are in good agreement with the previous studies. As shown in Fig. 11b, the computed drag coefficients also converge with first order as expected.

Table 2: Comparison of the computed drag coefficient and Strouhal number with previous experimental and numerical studies for the $Re = 100$ cylinder flow problem.

| Study | $\overline{C_D}$ | $St$ |
| --- | --- | --- |
| Present work ($D/\Delta x = 160$) | 1.41 | 0.165 |
| Relf [33] (experiment) | 1.39 | - |
| Wieselsberger [36] (experiment) | 1.40 | - |
| Williamson [37] (experiment) | - | 0.165 |
| Tseng and Ferziger [34] (simulation, resolution: $\pi D/\Delta x \approx 72$) | 1.42 | 0.165 |
| Lai and Peskin [37] (simulation, resolution: $D/\Delta x \approx 38$) | 1.45 | 0.165 |

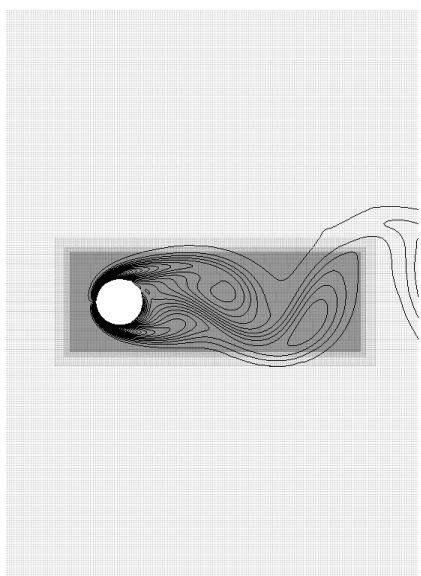

(a) Instantaneous vorticity contours overlaid on the mesh in the vicinity of the cylinder.

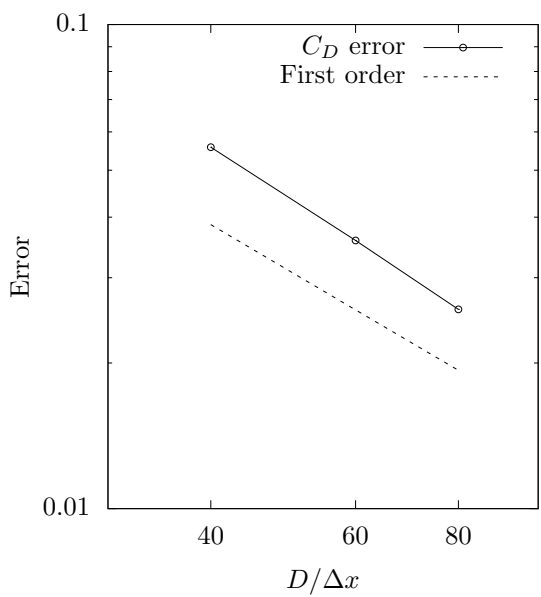

(b) Convergence plot of $C_D$ computation.

Figure 11: Results of vorticity and drag coefficient convergence for the $Re = 100$ cylinder flow problem.

### 5.4. Shock reflection from a wedge

For the compressible flow problem of the shock reflection from a wedge, we use the simulation parameters from Graves et al. [14] shown in Table 3. As illustrated in Fig. 12, the initial conditions are zero velocity with a discontinuity in pressure and density. The horizontal ground is also discretised with cut cells of volume fraction $10^{-3}$. A base resolution of $1024 \times 512$ cells is employed with two AMR levels of refinement factor 2 each to resolve the waves and cut cell interfaces. The minimum encountered cut cell volume fraction on the wedge surface is $4.9 \times 10^{-8}$. The ground and wedge are no-slip boundaries, while transmissive boundary conditions are specified at the left, right and top domain edges.

Table 3: Initial conditions (with reference to Fig. 12) for the shock reflection from wedge problem.

| Variable | Value |
| --- | --- |
| $p_0$ (Pa) | $1.95 \times 10^3$ |
| $p_1$ (Pa) | $7.42 \times 10^5$ |
| $\rho_0$ (kg/m$^3$) | $3.29 \times 10^{-2}$ |
| $\rho_1$ (kg/m$^3$) | $3.61 \times 10^{-1}$ |
| $\mu$ (kg/(m s)) | $1.21 \times 10^{-3}$ |
| $C_v$ (J/(kg K)) | $3.00 \times 10^2$ |
| $\xi$ (W/(m K)) | $1.7 \times 10^{-2}$ |
| $\gamma$ | 5/3 |

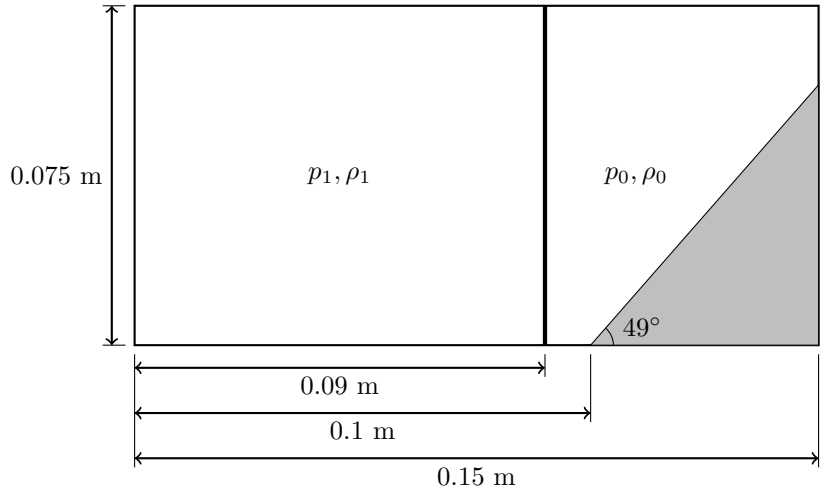

Figure 12: Simulation set-up for the shock reflection from wedge problem.

The solution evolves to form a left-propagating rarefaction and a right-propagating $M = 7.1$ shock. A separation bubble is created as the shock reflects off the ground boundary layer. Subsequent compression from the shock reflected off the wedge causes boundary layer reattachment. Fig. 13 shows density contour plots of the solution at $t = 9.6\mu$s. Note that the extra wave visible behind the shock in Fig. 13a is the numerical 'start-up error' [38] caused by starting the simulation with sharply discontinuous initial conditions.

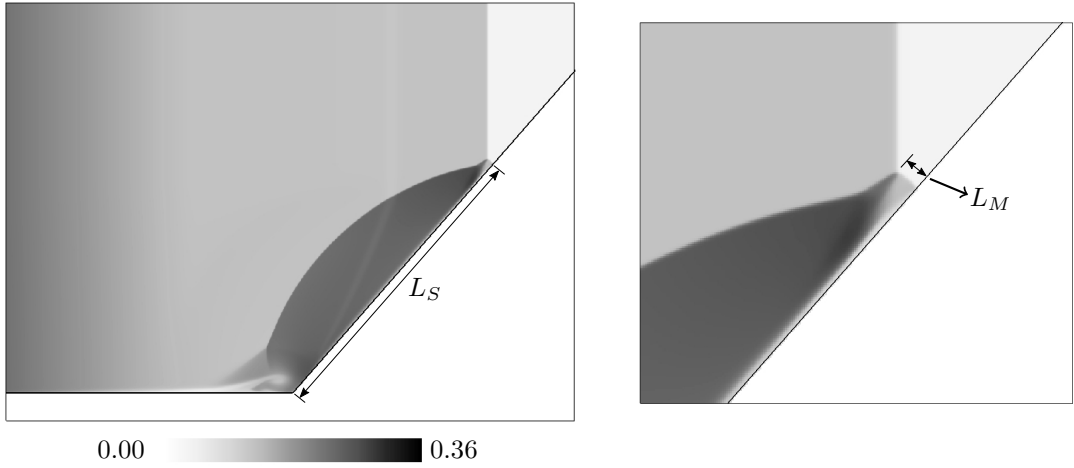

(a) $L_S$ is the distance travelled by the Mach stem up the wedge.

(b) Close-up of the Mach stem of length $L_M$.

Figure 13: Density contour plots at $t = 9.6\mu$s of the solution for the shock reflection from wedge problem.

The parameter of interest is

$$R_M = \frac{L_M}{L_S}, \qquad (26)$$

the ratio of the Mach stem length to the distance travelled by the Mach stem up the wedge. $L_S$ is determined from a line-out of the solution taken along the wedge surface. To avoid ambiguity in the determination of the position of the triple point, we compute $L_M$ via an algebraic approach. Consider Fig. 14, which illustrates the position of the incident shock (ET) and Mach stem (TD) at the end of the simulation. Once distances BD ($L_S$) and AC are known, TD ($L_M$) and hence, $R_M$ can be calculated via trignometry. $AC$ is determined from another line-out of the solution taken horizontally through the incident shock. Assuming an error of $\pm \Delta x$ in the determination of the average positions of the incident shock and Mach stem from the lineouts, we calculate $R_M = 0.0293 \pm 0.004$. Table 4 shows a comparison of our computed value for $R_M$ with previous numerical studies. The results from all three studies agree to 1 significant figure - differences in measurement procedures used in the studies are the most likely explanation for why closer agreement is not obtained. In our results, for example, $\Delta x / L_M \approx 6.3\%$, suggesting that the reported value for $R_M$ is very sensitive to the method used to measure $L_M$.

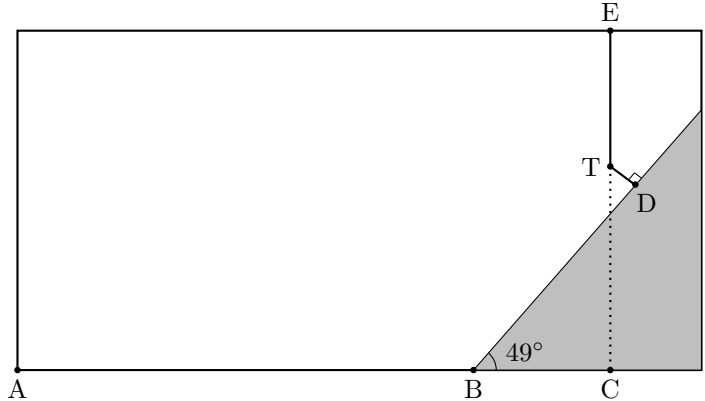

Figure 14: Illustration of the positions of the incident shock (ET) and Mach stem (TD) at the end of the shock reflection from wedge simulation.

Table 4: Comparison of the computed value of $R_M$ with previous numerical studies for the shock reflection from wedge problem.

| Study | $R_M$ |
|---|---|
| Present work | $0.0293 \pm 0.004$ |
| Graves et al. [14] | 0.03 |
| Al-Marouf and Samtaney [39] | 0.027 |

### 5.5. Three-dimensional supersonic flow over a sphere

The final test problem is that of three-dimensional, supersonic flow over a sphere for $M_\infty = 2.0$ and $Re_D = 6.5 \times 10^5$. The sphere of diameter $D = 0.1$ m is placed at a distance of $2.5D$ from the inlet in a domain of size $10D \times 5D \times 5D$. The free stream density $\rho_\infty$, velocity $u_\infty$ and temperature $T_\infty$ are set to be 0.2199 kg/m$^3$, 527.2 m/s and 173 K respectively as in Uddin et al. [40]. A supersonic inflow boundary condition is specified at the inlet, while transmissive boundary conditions are used at all the other domain boundaries. A base resolution of $100 \times 50 \times 50$ cells is employed and three levels of AMR of refinement factor 2 each are used to resolve the interface and sphere bow shock. The minimum encountered cut cell volume fraction is $2 \times 10^{-5}$. The simulation is run until a time of $t = 100D/u_\infty$. Fig. 15 shows a plot of the AMR patches overlaid on contours of normalised velocity magnitude ($|\mathbf{v}|/u_\infty$) along the $x$-$y$ symmetry plane of the sphere at the end of the simulation. Clearly visible in the figure are the bow shock ahead of the sphere and the viscous wake behind it.

The two results parameters of interest to us to validate the solution are the shock standoff distance $\delta/D$, and the drag coefficient $C_D$. $\delta$ is the shortest distance from the shock to the sphere leading edge. As seen in Table 5, our computed values for $\delta/D$ and $C_D$ show good agreement with previous experimental and numerical studies. Note that using one less AMR refinement factor produced no change to the drag coefficient computed to three significant figures, suggesting that our final resolution is sufficient to produce a numerically converged solution.

Table 5: Comparison of the computed shock standoff distance and drag coefficient with previous experimental and numerical studies for the three-dimensional supersonic flow over a sphere problem.

| Study | $\delta/D$ | $C_D$ |
|---|---|---|
| Present work | 0.175 | 1.01 |
| Krasil'shchikov and Podobin [41] (experiment) | 0.17 | 1.00 |
| Bailey and Hiatt [42] (experiment) | - | 1.00 |
| Uddin et al. [40] (simulation) | 0.18 | 1.07 |
| Al-Marouf and Samtaney [39] (simulation) | 0.183 | 0.96 |

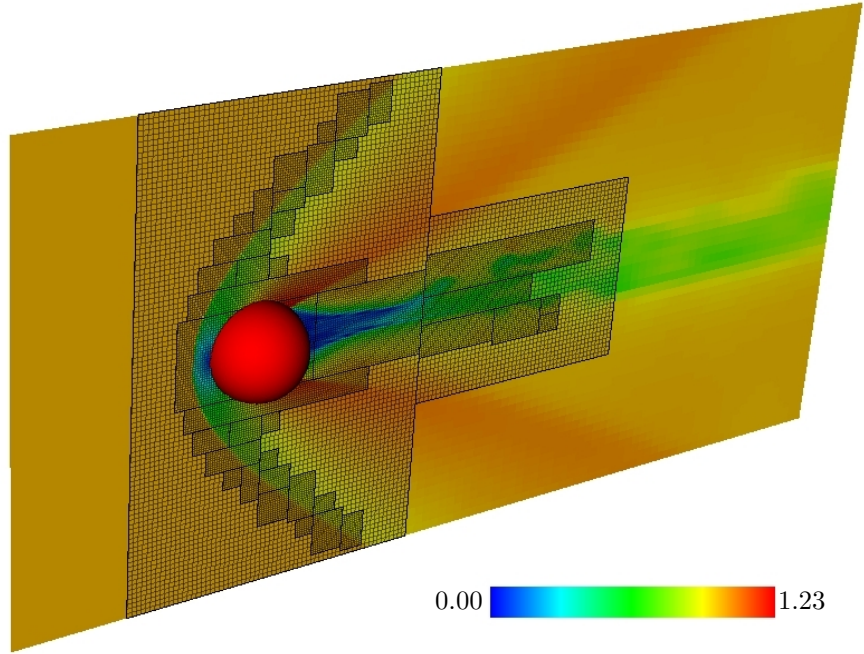

Figure 15: Plot of AMR patches on the second and third refinement level overlaid on a contour plot of $|\mathbf{v}|/u_\infty$.

## 6. Conclusions

In this paper, we presented a dimensionally split Cartesian cut cell method for the compressible Navier-Stokes equations. The numerical performance of the scheme was investigated through a number of test problems ranging from the nearly incompressible to the highly compressible flow regimes.

The computation of a nearly incompressible laminar boundary layer over a flat plate in both horizontal and inclined configurations was used to demonstrate that the method can handle coordinate aligned and non-aligned interfaces. The numerical convergence of the method was verified explicitly by computing solutions for a $Re = 20$ lid-driven cavity problem, and for flow over a circular cylinder at $Re = 40$ and $Re = 100$.

For the highly compressible problem of a $M = 7.1$ shock reflecting off a wedge, the complex shock boundary layer interaction pattern was accurately captured by the method, and the computed ratio of the length of the Mach stem to the distance travelled by it up the ramp showed good agreement with previous numerical investigations. Finally, the three-dimensional performance of the method was verified by computing a supersonic flow over a sphere at $Re = 6.5 \times 10^5$. The computed shock standoff distance and drag coefficient showed good agreement with previous experimental and numerical studies.

It is relatively straightforward to extend this work to be applicable for problems in the turbulent flow regime by adopting an explicit or implicit Large Eddy Simulation approach combined with a wall function to compute the wall shear stresses. We have made progress on implementing such a 'wall-modelled LES' scheme in the context of the split method and hope to present this as part of a future publication. For the future, modifying the method to achieve second-order accuracy at the boundary for both inviscid and viscous flows would also be a very useful contribution.

### Acknowledgements

Nandan Gokhale acknowledges the Cambridge Commonwealth, European & International Trust for financially supporting his research, and thanks Lukas Wutschitz for useful discussions. Rupert Klein acknowledges support by Deutsche Forschungsgemeinschaft through the Collaborative Research Centers CRC 1029 (project C01) and CRC 1114 (project C01). The authors would like to thank Philip Blakely for providing support with the parallel AMR code.

# References

# References

[1] R. Mittal, G. Iaccarino, Immersed boundary methods, Annu. Rev. Fluid Mech. 37 (2005) 239–261.
[2] M. Berger, Cut cells: Meshes and solvers, in: Handbook of Numerical Analysis, Vol. 18, Elsevier, 2017, pp. 1–22.
[3] D. K. Clarke, H. Hassan, M. Salas, Euler calculations for multielement airfoils using cartesian grids, AIAA journal 24 (3) (1986) 353–358.
[4] J. J. Quirk, An alternative to unstructured grids for computing gas dynamic flows around arbitrarily complex two-dimensional bodies, Computers & fluids 23 (1) (1994) 125–142.
[5] C. Helzel, M. J. Berger, R. J. LeVeque, A high-resolution rotated grid method for conservation laws with embedded geometries, SIAM Journal on Scientific Computing 26 (3) (2005) 785–809.
[6] M. Berger, C. Helzel, A simplified h-box method for embedded boundary grids, SIAM Journal on Scientific Computing 34 (2) (2012) A861–A888.
[7] P. Colella, D. T. Graves, B. J. Keen, D. Modiano, A Cartesian grid embedded boundary method for hyperbolic conservation laws, Journal of Computational Physics 211 (1) (2006) 347–366.
[8] X. Hu, B. Khoo, N. A. Adams, F. Huang, A conservative interface method for compressible flows, Journal of Computational Physics 219 (2) (2006) 553–578.
[9] D. Hartmann, M. Meinke, W. Schröder, A strictly conservative Cartesian cut-cell method for compressible viscous flows on adaptive grids, Computer Methods in Applied Mechanics and Engineering 200 (9) (2011) 1038–1052.
[10] L. Schneiders, D. Hartmann, M. Meinke, W. Schröder, An accurate moving boundary formulation in cut-cell methods, Journal of Computational Physics 235 (2013) 786–809.
[11] L. Schneiders, C. Günther, M. Meinke, W. Schröder, An efficient conservative cut-cell method for rigid bodies interacting with viscous compressible flows, Journal of Computational Physics 311 (2016) 62–86.
[12] M. Berger, M. J. Aftosmis, S. R. Allmaras, Progress Towards a Cartesian Cut-Cell Method for Viscous Compressible Flow, Aerospace Sciences Meetings, American Institute of Aeronautics and Astronautics, 2012.
[13] M. J. Berger, M. J. Aftosmis, An ODE-based Wall Model for Turbulent Flow Simulations, AIAA Journal (2017) 1–15.
[14] D. Graves, P. Colella, D. Modiano, J. Johnson, B. Sjogreen, X. Gao, A cartesian grid embedded boundary method for the compressible Navier–Stokes equations, Communications in Applied Mathematics and Computational Science 8 (1) (2013) 99–122.
[15] B. Muralidharan, S. Menon, A high-order adaptive Cartesian cut-cell method for simulation of compressible viscous flow over immersed bodies, Journal of Computational Physics 321 (2016) 342–368.
[16] N. Gokhale, N. Nikiforakis, R. Klein, A dimensionally split Cartesian cut cell method for hyperbolic conservation laws, Journal of Computational Physics 364 (2018) 186–208.
[17] R. Klein, K. Bates, N. Nikiforakis, Well-balanced compressible cut-cell simulation of atmospheric flow, Philosophical Transactions of the Royal Society of London A: Mathematical, Physical and Engineering Sciences 367 (1907) (2009) 4559–4575.
[18] M. J. Berger, J. Oliger, Adaptive mesh refinement for hyperbolic partial differential equations, Journal of Computational Physics 53 (3) (1984) 484–512.
[19] G. Strang, On the construction and comparison of difference schemes, SIAM Journal on Numerical Analysis 5 (3) (1968) 506–517.
[20] R. J. LeVeque, Finite Volume Methods for Hyperbolic Problems, Vol. 31, Cambridge University Press, 2002.
[21] Z. Dragojlovic, F. Najmabadi, M. Day, An embedded boundary method for viscous, conducting compressible flow, Journal of Computational Physics 216 (1) (2006) 37–51.
[22] E. F. Toro, Riemann solvers and numerical methods for fluid dynamics: a practical introduction, Springer Science & Business Media, 2013.
[23] S. Mauch, A fast algorithm for computing the closest point and distance transform, Tech. Rep. caltechASCI/2000.077, California Institute of Technology (2000).
[24] W. E. Lorensen, H. E. Cline, Marching cubes: A high resolution 3d surface construction algorithm, in: ACM siggraph computer graphics, Vol. 21, ACM, 1987, pp. 163–169.
[25] C. Günther, D. Hartmann, L. Schneiders, M. Meinke, W. Schröder, A cartesian cut-cell method for sharp moving boundaries, AIAA Paper 3387 (2011) 27–30.
[26] B. Paul, Calculating The Area And Centroid Of A Polygon), URL http://www.seas.upenn.edu/~sys502/extra_materials/Polygon%20Area%20and%20Centroid.pdf, Accessed: 16-02-2018.
[27] Z. Wang, Improved formulation for geometric properties of arbitrary polyhedra, AIAA journal 37 (10) (1999) 1326–1327.
[28] M. Meyer, A. Devesa, S. Hickel, X. Hu, N. Adams, A conservative immersed interface method for large-eddy simulation of incompressible flows, Journal of Computational Physics 229 (18) (2010) 6300–6317.
[29] F. Capizzano, Turbulent Wall Model for Immersed Boundary Methods, AIAA journal 49 (11) (2011) 2367–2381.
[30] Z. L. Chen, S. Hickel, A. Devesa, J. Berland, N. A. Adams, Wall modeling for implicit large-eddy simulation and immersed-interface methods, Theoretical and Computational Fluid Dynamics 28 (1) (2014) 1–21.